\documentclass[sigconf]{acmart}
\AtBeginDocument{%
  }

\setcopyright{cc}
\setcctype{by}
\renewcommand\footnotetextcopyrightpermission[1]{%
  \footnotetext{\textit{Presented at the workshop on Women’s Health as Cooperative Work: Participatory AI Across Care, Community, and Policy, held at Computer Supported Cooperative Work and Social Computing (CSCW) 2026, October 10–14, 2026, Salt Lake City, Utah, USA.}}
}\acmDOI{}
\acmISBN{}
  \acmConference[CSCW '26 Workshop]
  {Workshop on Women’s Health as Cooperative Work: Participatory AI Across Care, Community, and Policy}
  {October 10--14, 2026}
  {Salt Lake City, Utah, USA}
\begin{document}

\title{Making Doula Coordination Breakdowns Visible: \\A Participatory Case Study}


\author{Emma Ricci-De Lucca}
\correspondingauthor
% \email{emmariccid.th@dartmouth.edu}
\orcid{0000-0003-4150-5294}
\author{Sahithi Medikondla}
\author{Ashley Kim}
\author{Iha Rastogi}
\author{Elizabeth L. Murnane}
\affiliation{%
  \institution{Thayer School of Engineering\\ Dartmouth College}
  \city{Hanover}
  \state{New Hampshire}
  \country{USA}
}

\author{Saanvi Chilakapati}
\author{Jillian Berch}
\author{Karen L. Fortuna}
\affiliation{%
  \institution{Geisel School of Medicine\\ Dartmouth College}
  \city{Hanover}
  \state{New Hampshire}
  \country{USA}
}

\author{Margaret R. Emerson}
\affiliation{
  	\institution{University of Nebraska\\ Medical Center}
  \city{Omaha}
  \state{Nebraska}
  \country{USA}}

\renewcommand{\shortauthors}{Ricci-De Lucca et al.}

\begin{abstract}
  Doulas are trained, non-medical professionals who provide emotional, physical, and informational support to pregnant individuals before, during, and after childbirth. Doulas provide extensive relational and coordination labor for pregnant and postpartum patients, spending far more time with clients than traditional providers do, yet they operate with almost no doula-specific digital infrastructure. This gap is especially acute in rural communities, where postpartum depression affects about 25\% of women, well above the national average, and where geographic isolation and provider shortages leave many without access to behavioral health care. DoulaConnect began as a 12-week evidence-based postpartum texting curriculum, piloted separately in a randomized controlled trial in rural New England. Building on that work, we present DoulaConnect’s development into an AI-enabled platform co-designed with doulas and rural mothers through participatory human-centered design, as a case for surfacing where coordination breaks down in doula practice and who currently absorbs the resulting burden. Our community-based participatory study, conducted across the platform's Discover and Define phases, combines qualitative sessions with doulas and mothers to identify these breakdowns. We close with tensions this approach surfaces and open questions for the workshop about how these findings will inform future participatory AI infrastructure for doula care in rural and other under-resourced communities.

\end{abstract}

\begin{CCSXML}
<ccs2012>
   <concept>
       <concept_id>10003120.10003130.10011762</concept_id>
       <concept_desc>Human-centered computing~Empirical studies in collaborative and social computing</concept_desc>
       <concept_significance>300</concept_significance>
       </concept>
 </ccs2012>
\end{CCSXML}

\ccsdesc[300]{Human-centered computing~Empirical studies in collaborative and social computing}

\keywords{Maternal health, mental health, doula, digital tools, breakdowns}


% \raggedbottom

\maketitle

% \begin{figure}[h]
%   \centering
%  \includegraphics[width=\linewidth]{sample-franklin}
%   \caption{1907 Franklin Model D roadster. Photograph by Harris \&
%     Ewing, Inc. [Public domain], via Wikimedia
%     Commons. (\url{https://goo.gl/VLCRBB}).}
%   \Description{A woman and a girl in white dresses sit in an open car.}
% \end{figure}

% INSERT WORKSHOP BLOCK HERE
\noindent\textbf{Workshop:}
CSCW 2026 -- Women's Health as Cooperative Work:
Participatory AI Across Care, Community, and Policy
(October 11, 2026).

\par\smallskip
\noindent\textbf{Proceedings index:}
\url{https://arxiv.org/abs/2611.03776}

\par\smallskip
\noindent\textbf{Assigned report number:}
\texttt{CSCW-WHCW-2026-03}

\par\medskip

\section{Doulas for Maternal Mental Health Care}
\label{sec:introduction} % Assigning the label
Perinatal mood and anxiety disorders (PMADs), including postpartum depression, anxiety, and psychosis, are the most common complication of pregnancy and childbirth in the United States \cite{ayala2023perinatal, gavin2005perinatal}. With about one in eight women reporting symptoms of postpartum depression after giving birth in the U.S. \cite{cdc2022maternaldepression}, this figure stands as the highest rate of maternal mortality and morbidity of all high-income nations \cite{hoyert2020maternal}. Limited postpartum visits and inconsistent mental health screenings leave 75\% of those suffering from perinatal mental health conditions underdiagnosed or untreated \cite{cohen2021postpartum, garapati2023postpartum, griffen2021perinatal}. When left untreated, these conditions can lead to the development of long-term health conditions for the mother \cite{who2023lastinghealth}, stunted cognitive and social development for the child \cite{yan2016mothers}, and a cumulative estimated societal cost of \$14.2 billion per year in the U.S. alone \cite{luca2019societal}.

Doulas are increasingly recognized as an important part of the response to the perinatal mental health crisis. Doula support is extensively linked to improved maternal mental health outcomes \cite{horan2024doula}, yet the infrastructure available to support their day-to-day work remains limited. Existing digital health tools have been developed almost exclusively for patients or for clinicians, leaving doulas to coordinate an extensive portfolio of relational, administrative, and informational labor without any system built around their own workflow. Our own qualitative study found mHealth and digital health tools built for patients and for clinicians, but few designed with doulas as the primary end user \cite{chilakapati2025barriers}.

Rural communities face compounded risk that makes this issue more urgent. Postpartum depression rates among rural women range from approximately 16.7\% to 32.7\%, compared to a national average closer to 12\%, driven by geographic isolation, poverty, and limited access to behavioral health care \cite{policy2025addressing}. For women in these maternal mental health professional shortage areas, there is an urgent need for targeted interventions to expand maternal mental health providers and programs, such as postpartum doulas \cite{britt2023maternal}.

This paper presents DoulaConnect’s \textit{Discover} and \textit{Define} phases as a case of surfacing these breakdowns directly from doulas’ own accounts of their work, with the resulting participatory infrastructure (\textit{Develop} and \textit{Deploy} phases) still in progress.

\begin{figure}[h] % or [t]
  \centering
  \includegraphics[width=0.8\linewidth]{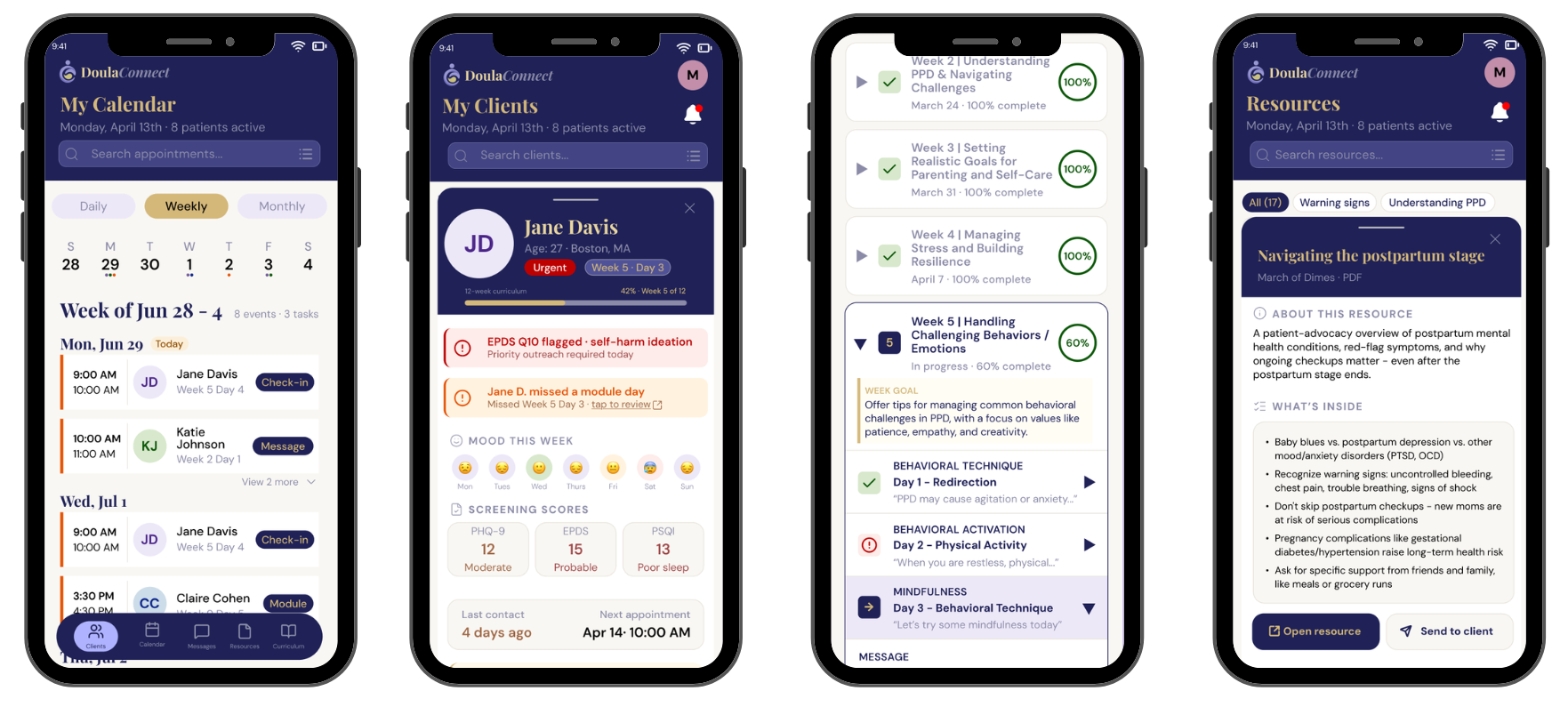}
  \caption{DoulaConnect prototype, showing the calendar for scheduling appointments, the client profiles, the curriculum for delivering the 12-week evidence-based texting program, and the resource library.}
  \Description{DoulaConnect Prototype}
  \label{fig:figmaprototype}
\end{figure}

\section{Surfacing Doula Coordination Breakdowns Through Participatory Methods}
\label{sec:methods}
\begin{figure}[h] % or [t]
  \centering
  \includegraphics[width=0.8\linewidth]{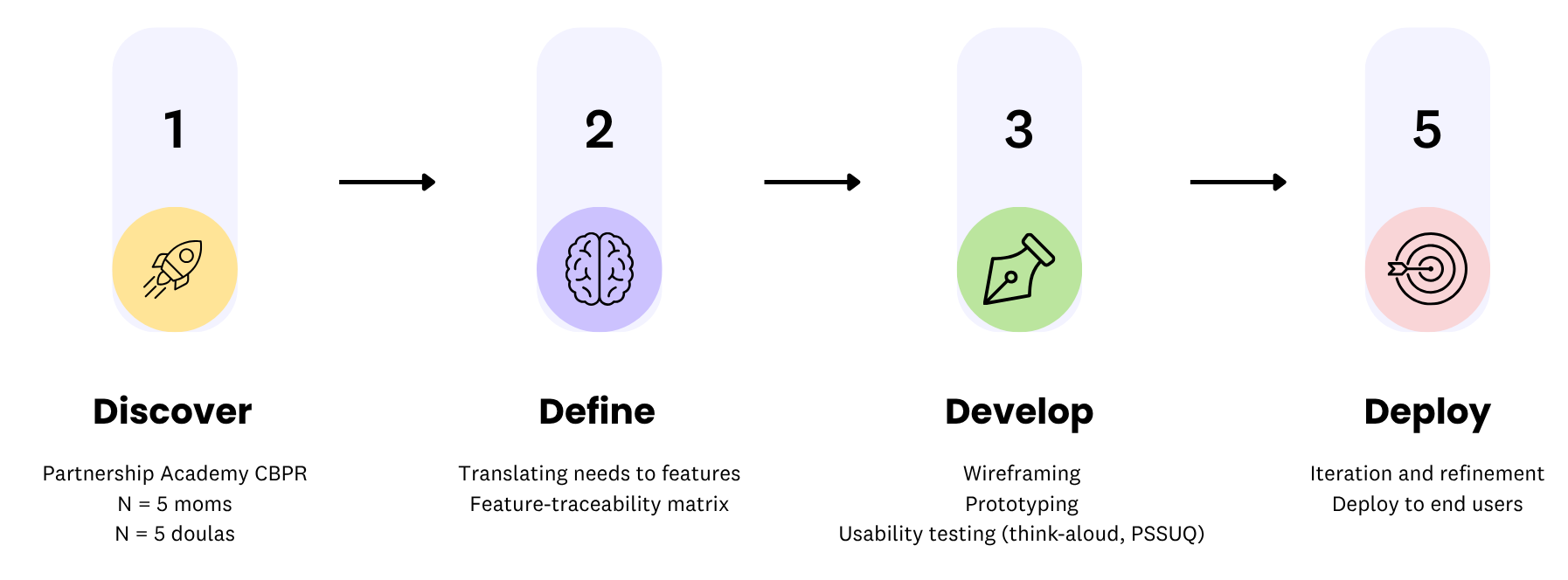}
  \caption{The human-centered design process guiding DoulaConnect's development, moving through four phases: Discover, Define, Develop, and Deliver.}
  \Description{Double Diamond Design Process}
  \label{fig:doublediamond}
\end{figure}

DoulaConnect is an AI-enabled, doula-facing platform developed through a human-centered design process \cite{designcouncil2005} that incorporates community-based participatory research (CBPR) \cite{collins2018community} across all of its phases (Figure \ref{fig:doublediamond}). The goal was to let the voice of the end user shape which workflow problems that platform addressed.

\subsection{Evidence-Based DoulaConnect Texting Curriculum}
\label{sec:curriculum}
DoulaConnect began as a text-based support program, utilizing a 12-week evidence-based curriculum where doulas provide daily mental, emotional, and social support to mothers experiencing postpartum depression (Figure \ref{fig:figmaprototype}). DoulaConnect has been developed using community-based participatory research (CBPR) with doulas, clinicians, researchers, and mothers to support postpartum mental health wellness. The curriculum provides doulas with evidence-based interventions that are designed to help mothers manage anxiety, build resilience, engage in values-driven parenting, and create support systems. Postpartum mothers receive simple and accessible texts from their doulas, grounded in clinical interventions such as psychoeducation, mindfulness, goal-setting, and relaxation techniques. The text-based format promotes flexibility and accessibility for mothers navigating the challenges of early parenting, especially those who may face barriers to traditional in-person care.

The 12-week evidence-based curriculum was pilot-tested in rural New England to understand the impacts on self-reported postpartum depression, parenting stress, parental self-efficacy, social support, and overall self-reported health. The pilot, randomized control trial examined feasibility, acceptability, and potential effectiveness of DoulaConnect (N = 15) compared to Treatment As Usual (N = 15) among adults aged 18+ with self-reported postpartum depression. In addition, semi-structured interviews were conducted pre and post implementation of the curriculum with key stakeholders like doulas, mothers, clinicians, and policy-makers to understand barriers and facilitators for scaling DoulaConnect in health systems beyond rural New England. As the pilot study is now concluded, preliminary results show the DoulaConnect group demonstrated greater reductions in Edinburgh Postnatal Depression Scale (EPDS) scores than the Treatment as Usual group.

% Utilizing both the qualitative and quantitative results from the pilot study, the text-messaging program will be integrated into an AI-enabled, doula-facing platform to equip doulas by providing evidence-based postpartum support and manage all their workflows in one digital platform. Qualitative evidence from stakeholders throughout the pilot revealed that doulas do not have a singular platform to oversee their client-base, track postpartum health outcomes, and employ their resources. 

\subsection{Discover}
\label{sec:discover}
We conducted virtual sessions with N=5 doulas and N=5 moms using the Partnership Academy framework, an experiential CBPR engagement tool for co-creation with stakeholders. The sessions collected doulas’ current workflows, frictions, and usage of digital tools, and mothers’ interactions with doulas and digital postpartum mental health support. Audio recordings and transcripts were analyzed using the Rigorous and Accelerated Data Reduction (RADaR) \cite{watkins2017rapid}  technique with member-checking. We identified six conceptual domains in doulas' postpartum workflow: friction with workflow management and administrative tasks, strengths of remote postpartum care, important metrics for the postpartum journey, strengths and limitations of current digital tools, opportunities for AI-assisted communication, and barriers to remote care.

\subsection{Define}
\label{sec:define}
These domains identified during discovery directly informed the generation of the use-case scenarios and candidate features considered for the creation of the platform. Each new feature was linked to specific interface elements through a feature-traceability matrix, so that every component could be tracked back to the community-identified need that produced it. Features were then ranked independently by the research team and by the doula team using a Likert scale, and the highest-ranked features were carried forward into the \textit{Develop} phase. Independent ranking by the research team and by doulas did not always converge, and where it diverged, that gap marked a point where researcher assumptions about priority did not match what doulas identified as most pressing in their own daily work. Ranking also surfaced disagreement among doulas themselves, a reminder that “the doula perspective” is not a single, uniform stakeholder position but one negotiated across differing caseloads, agencies, and working styles. The lesson we would share with the workshop participants is that a multi-stakeholder ranking exercise does more than sort a feature list. It exposes where priorities genuinely compete, both between researchers and the stakeholders they are designing with, and within that group of stakeholders itself, which outsiders too often treat as a single, uniform voice.

\subsection{Develop}
\label{sec:develop}
Resolving these competing priorities called for a flexible, adaptive approach rather than averaging out all the scores across different groups. For the first iteration of the DoulaConnect app design, we included every feature ranked highly by both the research team and the doulas, given agreement across independent rankings gave us the strongest confidence that a feature addressed a shared priority. We then included the next highest-ranked features that doulas alone had prioritized, and separately the next highest-ranked features that the research team had prioritized, in each case drawing roughly the top 25\% of each group's remaining list. The lower-ranked priorities that did not make this cut were not discarded. They are planned for later iterations, as no single release could accommodate every priority surfaced across both rankings given time and resource constraints. This also reflects our view that the rankings themselves are not final. As we continue to develop DoulaConnect, we expect ongoing participatory input to refine the platform and revise which priorities matter most and to surface others we may not have yet captured.

A second round of interviews will be conducted for doulas to conduct usability testing with the resulting prototypes through think-aloud structured tasks, open-ended questions, and standardized usability instruments such as the Post-Study System Usability Questionnaire (PSSUQ). Doulas complete a structured set of think-aloud tasks covering core workflows, including messaging a client, sharing a resource, customizing which client metrics are tracked, configuring automated reminders, and scheduling an appointment, after which they answer structured follow-up questions probing comprehension of the platform, trust and comfort in using it, and their emotional response to the experience. Our aim is to ensure that usability testing is a continued, iterative part of the participatory design process rather than a separate validation stage. Doulas are evaluating whether the tool matches the coordination problems they themselves named in the Discover phase.

\subsection{Deploy}
\label{sec:deploy}
This phase has not yet begun. Findings from the develop phase usability sessions will inform the refinements to DoulaConnect ahead of broader deployment. The changes will be made in response to what doulas themselves identified as priority features for administrative workflow support, remote postpartum monitoring, and AI-assisted communication from the prototype.

This process illustrates what it means to make coordination breakdowns visible through participation rather than assume them in advance. The features prioritized for development, administrative workflow support, remote postpartum monitoring, and AI-assisted communication, all trace back to specific frictions doulas named themselves, giving a concrete account of where their coordination work currently breaks down and who is left to repair it.

\section{Tensions in Building Participatory AI Infrastructure}
\label{sec:tensions}
Designing infrastructure \textit{alongside} a community rather than \textit{for} one does not eliminate difficult design choices. Sociotechnical systems research has argued for decades that infrastructure designed without ongoing input from the people who operate it tends to lead to this kind of friction, because no design process can fully anticipate every context of use in advance \cite{ruhleder1996steps}. Participatory methods build a feedback loop directly into the design process, in which breakdowns and revisions are an expected part of a system’s lifecycle \cite{ruhleder1996steps}. This approach surfaces difficult design choices earlier and in a form that a research team can act on.

% First, community-based doulas already report experiencing high stress and burnout \cite{guenther2022birthdoula}, so a new AI-enabled system carries the risk of becoming another demand on their limited time. Participatory input can shape what gets built, but it cannot guarantee that what is built will actually reduce burden once it enters daily use.

DoulaConnect will be deployed across different doula agencies, each of which brings its own established practices, caseload structures, and existing tools into the design process. A platform shaped around the workflow of one agency may not transfer cleanly to another, and features ranked as essential by one group of doulas may be redundant or poorly suited to a different agency’s circumstances. Scaling participatory infrastructure across agencies with distinct working conditions raises a question this paper does not resolve: how a shared platform can adapt to each agency's particular infrastructure and caseload norms without either fragmenting into agency-specific versions or flattening the differences that make each agency's practice distinct.

The ranking process described in Sections \ref{sec:define} and \ref{sec:develop} that was utilized to resolve competing priorities results in features that are ranked lower for both the research team and the doula participants, which are deferred to later iterations. If a minority of individual doulas consistently rank a given feature lower than the majority, participatory ranking alone will not protect that minority's priorities from being pushed downstream indefinitely.

As DoulaConnect begins incorporating features like automated reminders and metric-based alerts (e.g., mental health survey score, physical activity, breastfeeding), the platform starts taking on part of the coordination work that previously fell entirely on doulas. This raises the question of who is accountable when an automated action may malfunction, whether a doula can override or dismiss it, and who is responsible for follow-up if something automated goes unaddressed. Co-design and usability testing can surface doulas’ reactions to these features, but it cannot by itself settle where accountability should sit.

\section{What We Bring to the Workshop}
We see DoulaConnect’s \textit{Discover} and \textit{Define} phases as one case of using participatory methods to surface where coordination breaks down in an under-infrastructured area of care, and who currently performs the repair work when it does. We are interested in the workshop’s collaborative mapping and synthesis in other parts of the women’s health ecosystem, and to ask whether the same participatory methods that surfaced these gaps would surface similar patterns elsewhere. DoulaConnect illustrates how participatory design can reveal invisible coordination work before building AI infrastructure, offering one model for designing AI around care work rather than asking care workers to adapt to AI.

\section{Acknowledgments}
This material is based upon work supported in part by an award from Dartmouth Innovations Accelerator for Digital Health, the National Science Foundation under the award DGE- 2125733, and the Dartmouth College E. E. Just Program.

% \begin{acks}
% \end{acks}

% \section*{Ethics and Privacy Statement}

\bibliographystyle{ACM-Reference-Format}
\bibliography{sample-base}

% \appendix

\end{document}